\documentclass{rnaastex}

\shorttitle{Vortex trapping and backreaction}
\shortauthors{Lyra et al.}

\renewcommand{\fig}[1]{Fig.~\ref{#1}}

\begin{document}

\title{Pebble trapping backreaction does not destroy vortices}

\correspondingauthor{Wladimir Lyra}
\email{wlyra@csun.edu, wlyra@jpl.nasa.gov}

\author{Wladimir Lyra}
\affiliation{Department of Physics and Astronomy, California State
  University Northrige, 18111 Nordhoff St, Northridge CA 91130, USA}
\affiliation{Jet Propulsion Laboratory, California Institute of Technology, 4800 Oak Grove Drive, Pasadena, CA, 91109, USA}

\author{Natalie Raettig}
\affiliation{Max-Planck-Institut f\"ur  Astronomie, K\"onigstuhl 17, 69117 Heidelberg, Germany}

\author{Hubert Klahr}
\affiliation{Max-Planck-Institut f\"ur Astronomie, K\"onigstuhl 17, 69117 Heidelberg, Germany}

\maketitle

The formation of planets remains one of the most challenging problems
of contemporary astrophysics. Starting with micron-sized dust grains,
coagulation models \citep{Zsom+10} predict growth up to centimeter
size (hereafter called ``pebbles'') by electromagnetic hit-and-stick mechanisms. However, growth beyond this size is halted because fragmentation becomes increasingly more likely \citep{Benz00} and because the pebbles rapidly drift toward the star \citep{Weidenschilling77a,Brauer+08}. Ways to bypass this last problem have focused on inhomogeneities in the flow, in order to trap pebbles in their drift. Large grains and pebbles may be trapped in zonal flows \citep{Lyra+08a,Johansen+09,Simon+12} that are local inversions in the angular velocity profile, brought about by spatial variations in magnetic pressure. The pebbles themselves can give rise to the necessary inhomogeneities, as their drift is dynamically unstable and develops into the streaming instability \citep{YoudinGoodman05}, leading to intense pebble clumping \citep{JohansenYoudin07} and subsequent planetesimal formation \citep{Johansen+07}. Once a planetesimal forms, rapid accretion of pebbles ensues, as the aerodynamical drag force directs pebbles inside the planetesimal's Hill sphere to its center \citep{OrmelKlahr10,LambrechtsJohansen12}. 

Yet another formation mode is formation through vortices. Vortices are exact equilibrium solutions to the compressible Euler equation with barotropic equation of state, characterized as flows with closed elliptic streamlines. The study of vortices has been an active area of research in theory of circumstellar disks since \citet{BargeSommeria95}, \cite{AdamsWatkins95} and \cite{Tanga+96} independently proposed them as fast routes to planet formation. Because vortices are in equilibrium between the Coriolis and the pressure force, the pressureless grains will orbit along a vortex streamline experiencing a drag force. Similarly to the pebble drift in the disk, vortex-trapped pebbles will spiral inwards as well. Yet, whereas in the disk the solids drift to the inhospitable flames of the star, in a vortex they drift to the vortex eye, which has the convenient side effect of dramatically enhancing the solids-to-gas ratio locally. This is a very effective mechanism to concentrate pebbles \citep{Chavanis00,KlahrBodenheimer06}, as also seen in numerical simulations \citep{GodonLivio00,Johansen+04,FromangNelson05,InabaBarge06}, and possibly in ALMA observations \citep{vanderMarel+13}.

We showed in global 2D simulations with vortices that the concentration of pebbles easily reaches the conditions necessary to gravitationally collapse them into planets \citep{Lyra+08a, Lyra+09a, Lyra+09b,Meheut+12}. However, the scale height of the pebbles and of the gas are different, and in 3D the pebbles should settle toward the midplane forming a thin layer with a high pebble load. This is potentially dangerous, because the back-reaction of the drag force may disrupt the vortex column. Indeed, \cite{Raettig+15} show in high resolution local 2D simulations that a high pebble load destroys the vortex. Here we want to know if the same happens in three dimensions. 

We perform three-dimensional simulations with pebbles in a local box. The gas is unstratified 
but the pebbles feel the stellar vertical gravity and are allowed to sediment. This is because the 
scale height of the gas and the pebbles are very different, with the gas being essentially unstratified 
at the scales where the dust scale height is resolved. The gas is baroclinic, allowing the growth of 
the convective overstability, leading to a large-scale vortex. The
shearing box model used is that of \cite{LyraKlahr11}. The reader is
refered to that paper for the equations of motion. The model includes
the linearized pressure gradient, at the expense of dropping an
$x$-dependent term in the pdV work to keep shear-periodicity.   The simulation has mesh resolution of $N_x,N_y,N_z$=288$\times$288$\times$144. While the gas will be treated on a mesh, the pebbles will be treated numerically as particles. Gas and pebbles communicate via drag forces. Seven million Lagrangian particles represent the solids. We first evolve a box with only gas, for 400 orbits. Next we included the particles and evolve for further 150 orbits. We use particles of Stokes number St=0.05, equivalent to 5cm at 1AU and 0.5mm at 100AU. 

We solve the equations with the {\sc Pencil Code}. The code,
including improvements done for the present work, is publicly
available under a GNU open source license and can be downloaded at
\url{http://www.nordita.org/software/pencil-code}. 

\begin{figure*}
  \begin{center}
    \resizebox{\textwidth}{!}{\includegraphics{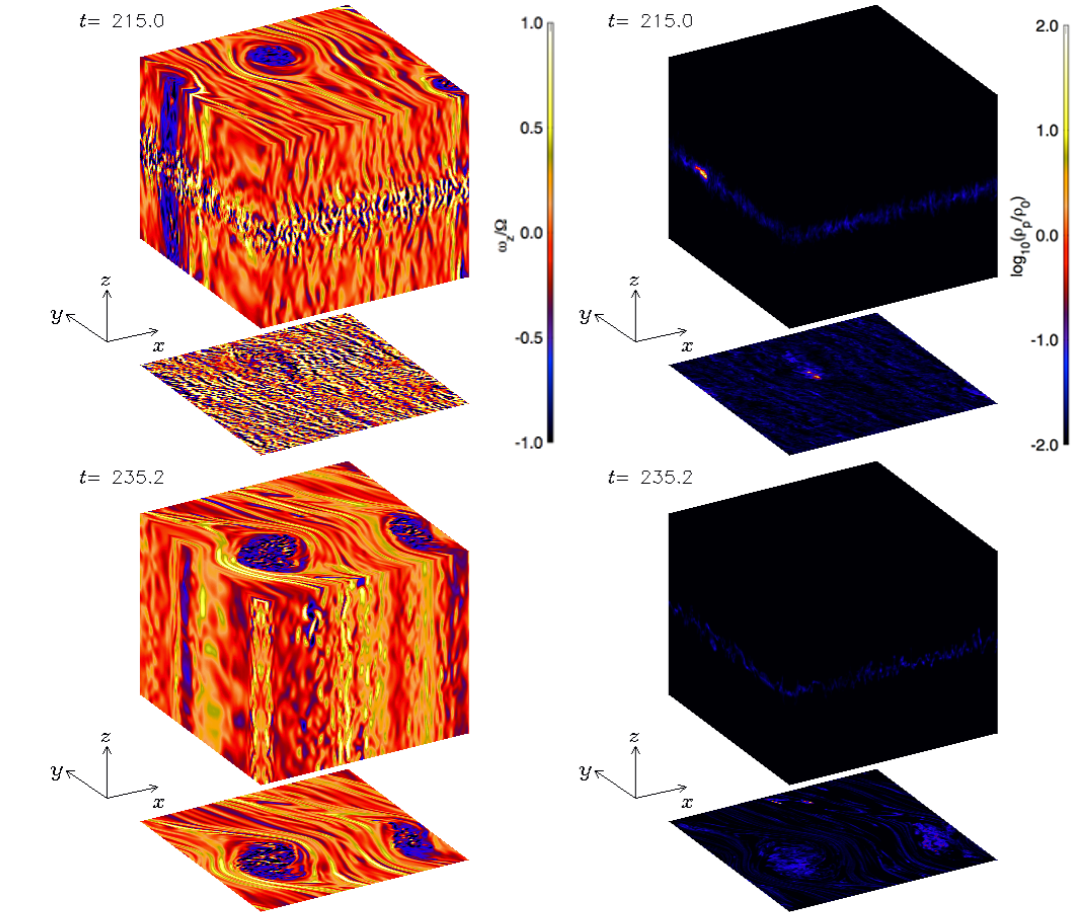}}
\end{center}
\caption{
{\it Upper panels:} Vorticity (left) and bulk density of
  solids (right) in a model with drag force backreaction. The
  particles concentrate in a thin layer around the midplane in the
  vortex core. Yet, the heavy particle load does not destroy the
  vortex column, that keeps concentrating particles. {\it Lower
    panels:} control model without backreaction.
}
\label{fig:3dvortex}
\end{figure*}

The results are shown in \fig{fig:3dvortex}. The left and right panels show the vorticity of the gas and the bulk density of pebbles, respectively. The upper panels show the model with backreaction. The lower panels represent a control run without backreaction. The lower plot is each frame is a slice through the midplane. Before the insertion of pebbles the box has saturated turbulence with convective overstability, with $\alpha \approx 5\times 10^{-3}$. Two large scale vortices are seen.  In the control runs without backreaction, the trapping of pebbles leads to a density enhancement of about a factor 30, as expected as a result of equilibrium between drag force and diffusion \citep{LyraLin13}. In the simulation with backreaction, it is seen that the pebbles disturb the vortex around the midplane, where the pebbles sediment. If all data we had was the midplane, i.e., if the simulation was 2D, we would conclude that the pebble load destroyed the vortex. Yet, as clear from the 3D box, although the pebbles do disturb the vortex around the midplane, the column does not get destroyed. This result is important because based on the previous 2D result suggesting complete disruption, the vortex interpretation of ALMA observations has been called into question. We show instead that the vortex behaves like a Taylor column, and the pebbles as obstacles to the flow. Pebble accumulation in the center of the vortices proceeds to roughly the same concentration as in the control run without backreaction. 

\acknowledgements

W. L. acknowledges support of Space
Telescope Science Institute through grant HST-AR-14572 and the NASA
Exoplanet Research Program through grant 16-XRP16\_2-0065. 
The simulations presented in this paper utilized the Stampede
cluster of the Texas Advanced Computing Center (TACC) through XSEDE grant TG-AST140014. 
This work was performed in part at the Jet Propulsion Laboratory, California Institute of Technology.

\bibliographystyle{apj}

\end{document}